# Silicon Photonic 2.5D Interposer Networks for Overcoming Communication Bottlenecks in Scale-out Machine Learning Hardware Accelerators

Febin Sunny, Ebadollah Taheri, Mahdi Nikdast, Sudeep Pasricha
*Department of Electrical and Computer Engineering*
Colorado State University
Fort Collins, Colorado, USA
{febin.sunny, ebad.taheri, mahdi.nikdast, sudeep}@colostate.edu

*Abstract*—Modern machine learning (ML) applications are becoming increasingly complex and monolithic (single chip) accelerator architectures cannot keep up with their energy efficiency and throughput demands. Even though modern digital electronic accelerators are gradually adopting 2.5D architectures with multiple smaller chiplets to improve scalability, they face fundamental limitations due to a reliance on slow metallic interconnects. This paper outlines how optical communication and computation can be leveraged in 2.5D platforms to realize energy-efficient and high throughput 2.5D ML accelerator architectures.

## I. INTRODUCTION

As modern machine learning (ML) applications scale in terms of memory use, communication bandwidth, and computational requirements at an unprecedented rate, system-level solutions to address these requirements are becoming a necessity. Larger ML hardware accelerator chips capable of higher computational throughput are emerging to meet these needs [1]. However, incorporating such computing capacity on a monolithic, single-chip architecture is difficult [2]. The challenges range from power and thermal restrictions to low fabrication yield [3]. As a result, modern accelerator architectures are moving towards 2.5D architectures, where multiple smaller chiplets are connected over an interposer, enabling high bandwidth communication and high throughput ML acceleration through co-packaging memory and processing.

2.5D integration has already found success in commercial accelerators and GPUs [4]. However, with emerging ML models such as transformers (used in large language models (LLMs)) becoming increasingly complex, these 2.5D platforms need to support very high bandwidths between chiplets. While state-of-the-art electrical wires on interposers can offer bandwidths approaching hundreds of Gb/s with an energy-efficiency of a few pJ/bit [5], there is a need to scale beyond 10 Tb/s bandwidth and fJ/bit energy efficiency, to sustain ML-related data transfer demands between chiplets. However, attenuation and inter-symbol interference from dispersion become significant issues at higher frequencies in electrical wires, limiting cutoff rates to 40 Gb/s [6]. Thus, electrical wires cannot meet the data transfer needs of emerging chiplet platforms for ML acceleration.

Silicon photonic (SiPh) interconnects can overcome the high energy consumption, limited bandwidth, and high latency of metallic interconnects [7]. SiPh links have many advantages in 2.5D platforms, including minimal signal attenuation, high bandwidth, low energy consumption, and the ability to leverage the mature CMOS ecosystem for low-cost fabrication. Further, ML workloads exhibit broadcast and multicast communication patterns [8] which can be efficiently implementable using networks of SiPh links [9]. Photonic devices can also be used to perform energy-efficient and high throughput ML computational operations, e.g., matrix multiplications [10].

In this paper, we highlight two innovations to enable scale-out hardware acceleration of ML workloads that benefit from SiPh in 2.5D chiplet platforms. The first contribution, TRINE (first discussed in [11]), is a novel 2.5D SiPh interposer network designed to efficiently connect electronic chiplets executing ML workloads. The second contribution, 2.5D-CrossLight (first discussed in [12]) extends the scope by utilizing SiPh for both communication and computation in 2.5D chiplet platforms.

## II. RELATED WORK

SiPh interposer networks are a promising communication substrate in emerging 2.5D platforms [13]. A few recent efforts, e.g., SPRINT [14] and SPACX [15] have proposed SiPh interposer networks designed specifically for handling traffic for ML workloads. The architecture of these networks is based on a bus communication model, where one or several writers send data to one or several readers using an optical waveguide. But the use of multiple writers/readers on the same waveguide increases the accumulated optical power losses exponentially. Compensating for these losses leads to increased laser power consumption overheads in these works. Our proposed TRINE SiPh interposer network architecture [11], discussed in Section IV, employs broadband switch devices arranged in a tree topology to mitigate losses associated with bus-based networks.

Several recent efforts have also proposed SiPh-based ML accelerators where computation is performed optically. CrossLight [16] was the first cross-layer optimized convolution neural network (CNN) accelerator. It made use of multiple wavelengths to perform computations in parallel for ML model inference. Operations with model parameters (e.g., multiplications of vectors of weights) were performed by imprinting parameters onto optical signal amplitudes using wavelength-selective devices, such as MRs (discussed in the next section). This accelerator was further extended to support transformer neural networks [17], recurrent neural networks [18], and graph neural networks [19]. However, these accelerators are designed on monolithic chips that have limited scalability. Our proposed 2.5D-CrossLight architecture [12], discussed in Section V, extends the scope of these SiPh accelerators for scale-out ML inference on 2.5D platforms.

### III. OVERVIEW OF SILICON PHOTONICS

There exist a variety of SiPh devices, and many ongoing research efforts are aiming to enhance their performance and efficiencies [20]. One prominent SiPh device widely employed in both communication and computation is the Microring Resonator (MR) which is a resonant device that can manipulate light of a specific wavelength (called its resonant wavelength). For example, an MR is depicted in figure 1(a) which is resonant with the red wavelength optical signal. During communication, an incoming optical signal to the *In* port can be dynamically switched between the *Through* and *Drop* ports according to the wavelength of the signal and the resonance state of the MR, which is tunable. The MR can operate as a modulator (figure 1(b)) where it encodes (sends) a '0' by absorbing the red signal, while allowing other signals (e.g., green) to pass unaltered. It can act as filter (figure 1(c)) where the MR is tuned to couple out the red signal from the *In* port to the *Drop* port while allowing other signals (e.g., green) to pass through unaltered. Further, by adjusting the fraction of the MR transmission from the *In* port to the *Drop* port, a weight of a neural network can be imprinted onto an optical signal.

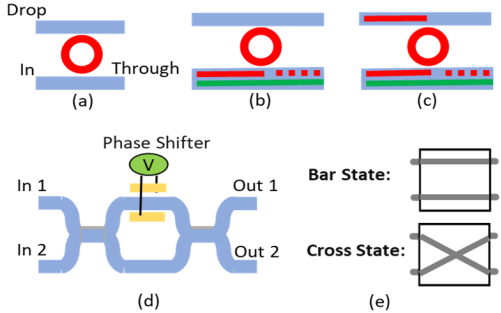

Figure 1: SiPh switching devices: (a) Microring resonator (MR), (b) MR modulator, (c) MR filter, (d) MZI switch, (e) MZI switch states.

The Mach-Zehnder Interferometer (MZI), shown in figure 1(d), is a SiPh device employed for switching when a broad-spectrum switching capability is required e.g., when there is a need to simultaneously switch multiple wavelengths from one port to another (figure 1(e)). Compared to the MR, the MZI has a much larger footprint and exhibits slower switching speeds.

Figure 2 illustrates a SiPh communication setup between a writer chiplet and a reader chiplet on a 2.5D platform. A laser is employed to generate three optical signals with distinct wavelengths, depicted in red, blue, and green. Each modulator on the writer side of the interposer modulates data onto its respective optical signal. On the reader side, the signals are filtered and subsequently converted to electrical signals using photodiodes. There are also several Multiply and Accumulate (MAC) processing elements (PEs) on each chiplet, which are connected to a gateway. Gateways are responsible for facilitating communication to the memory chiplet, which includes a global buffer (GLB), through the optical interposer.

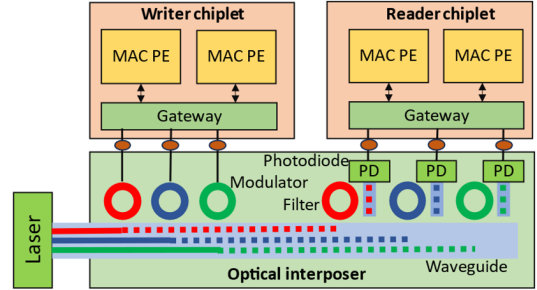

Figure 2: Silicon photonic communication to send data from a writer chiplet to a reader chiplet on a 2.5DF platform.

### IV. TRINE: PHOTONIC INTERPOSER NETWORK

Bus-based SiPh communication architectures in 2.5D platforms, as proposed in SPRINT [14] and SPACX [15] and summarized in figure 3(a), are energy inefficient due to the high-power consumption of the laser as the system scales up. This is attributed to the fact that as an optical signal passes through numerous MR filters/modulators, it experiences losses due to MRs' non-ideal frequency response. Consequently, the laser power must be increased to compensate for these losses, ensuring that the photodiodes at the reader side can receive a detectable signal, thereby enabling reliable communication.

To overcome the high-power consumption of the laser, a switch-based architecture can be employed, as illustrated in figure 3(b). A simple approach is to utilize a tree network to support one-to-many and many-to-one communication between the compute chiplets (i.e., chiplet with MAC PEs) and the memory chiplets. However, the memory bandwidth of the tree topology is restricted to one waveguide's bandwidth in this scenario, resulting in high latencies. TRINE [11] addresses this limitation by incorporating multiple tree subnetworks (figure 3(c)). The number of subnetworks can be tailored to match the bandwidth that the memory can provide, ensuring that the network bandwidth of memory aligns with the memory bandwidth. This approach maximizes performance without wasting network resources. Furthermore, TRINE reduces the number of network stages compared to a tree network, thereby minimizing optical losses and further saving energy.

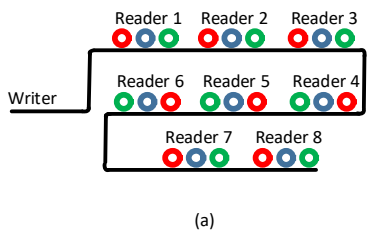 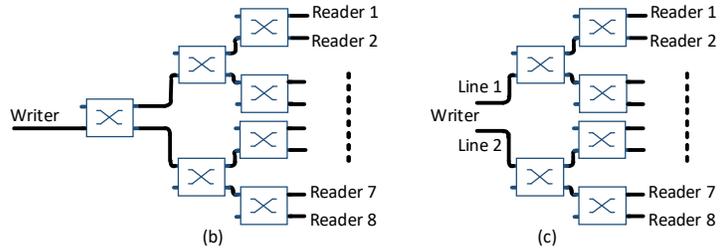

Figure 3: (a) Bus-based 2.5D interposer network used in SPRINT and SPACX, (b) Tree network, and (c) TRINE network.

We evaluated the TRINE network architecture against state-of-the-art SiPh interposer networks, namely SPACX [15] and SPRINT [14]. The maximum chiplet-interposer bandwidth was set at 100 GBs/chiplet, following [11], with limitations imposed by microbump density for vertical connections. With a modulation frequency of 12 GHz and a gateway frequency of 2 GHz, we opted for 8 subnetworks to use the maximum bandwidth offered by memory chiplets. Our evaluation encompassed six CNN models (DenseNet, ResNet, LeNet, VGG, MobileNet, and EfficientNet), with results summarized in Figure 4, comparing interposer network power, energy, and latency. All results are normalized to the results for the SPRINT network. In terms of power consumption, TRINE exhibits increased laser power usage compared to SPACX and Tree networks, due to overheads from its multiple subnetworks. Despite higher trimming power consumption compared to SPACX and Tree, TRINE shows notable improvements in latency and energy efficiency, as depicted in Figure 4. The use of 8 subnetworks and 32 gateways results in 2 switch stages for TRINE, contrasting with 5 stages in the Tree network topology. The reduced number of stages reduces the switch latency and area overhead and mitigates power losses associated with passing through MZIs in each stage, leading to lower latency and energy costs compared to the Tree and SPACX networks.

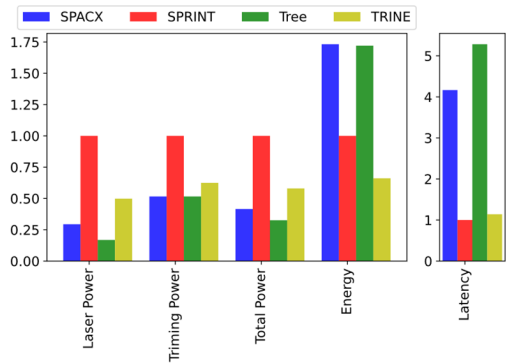

Figure 4: Performance evaluation of TRINE versus SPACX, SPRINT, and Tree network architectures.

## V. 2.5D-CROSSLIGHT ACCELERATOR ARCHITECTURE

As recent accelerators have shown, it is possible to use the same SiPh devices that are employed in the design of photonic networks (such as TRINE) for performing computations. Monolithic ML accelerators using SiPh have been designed [16]-[19], as discussed in Section II, but they face challenges with scalability and energy efficiency as ML model complexities grow. In [12] we proposed a SiPh-based 2.5D ML accelerator architecture called 2.5D-CrossLight that used SiPh components not just for photonic communication over the interposer network, but also photonic computation in the PEs within chiplets. This accelerator extended the CrossLight [16] photonic neural network accelerator, which was engineered for rapid execution of multiply and accumulate (MAC) operations using photonics, to the more scalable 2.5D chiplet platform.

Figure 5 shows a high-level representation of the chiplet- and SiPh-based 2.5D-CrossLight ML accelerator. The photonic MAC units in this design perform multiply operations using noncoherent photonics and uses balanced photodetectors for summing up partial products. Weights and activations are encoded onto wavelengths through wavelength-specific MR filters, following the broadcast-and-weight protocol [10]. The architecture features heterogeneous MAC unit sizes across chiplets, addressing various CNN convolution layer kernel sizes (e.g., 3×3 convolution MACs in Chiplet 1, 7×7 in Chiplet 2) and also larger-scale operations for fully connected layers.

The 2.5D-CrossLight ML accelerator also integrates a reconfigurable photonic 2.5D interposer network that can adapt inter-chiplet bandwidth based on real-time traffic requirements. By performing intelligent traffic load monitoring on electro-photonic gateways, the network adaptively activates or deactivates gateways using phase-change material couplers (PCMCs). A PCMC can tune the optical input of each writer dynamically, which enables opportunities for reducing laser power and thereby improving energy consumption of the interposer network. Additionally, deactivated gateways are power gated to further reduce energy consumption.

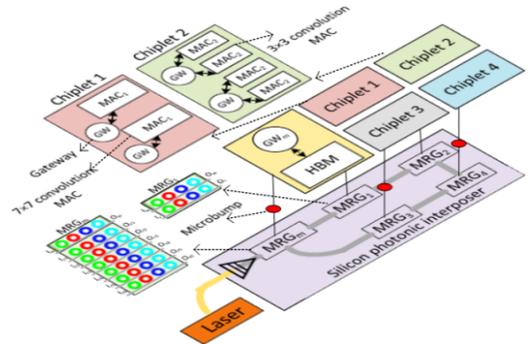

Fig. 5: Overview of proposed 2.5D interposer chiplet-based DNN accelerator architecture. From [12].

Inter-chiplet communication in 2.5D-CrossLight involves two types of traffic: 1) reading weights and inputs for MACs from memory, and 2) writing MAC outputs back to memory. For reads from memory to compute chiplets, we use the Single-Writer-Multiple-Readers (SWMR) protocol, while for writes from compute chiplets to memory, we employ the Single-Writer-Single-Reader (SWSR) protocol [20]. Consequently, a memory chiplets' Microring Resonator Group (MRG; figure 5) need multiple sets of MR filters to receive data from compute chiplets. Each compute chiplet requires only one set of MR filters for receiving data from memory. Both types of chiplets need a set of MR modulators for data transmission.

To analyze the 2.5D-CrossLight architecture, we performed several experiments with various ML workloads and determined the power dissipation, latency, and energy-per-bit. We considered two variants of the 2.5D-CrossLight architecture: one with a photonic interposer network as discussed above (called *2.5D-CrossLight-SiPh-Interposer*) and another with an electrical mesh interposer network from [21] (called *2.5D-CrossLight-Elec-Interposer*). These variants were intended to showcase the differences between using a photonic and electronic interposer network. We also compared these variants against the monolithic *CrossLight* SiPh-based ML accelerator, to showcase the differences between monolithic and scale-out 2.5D chiplet platform implementations.

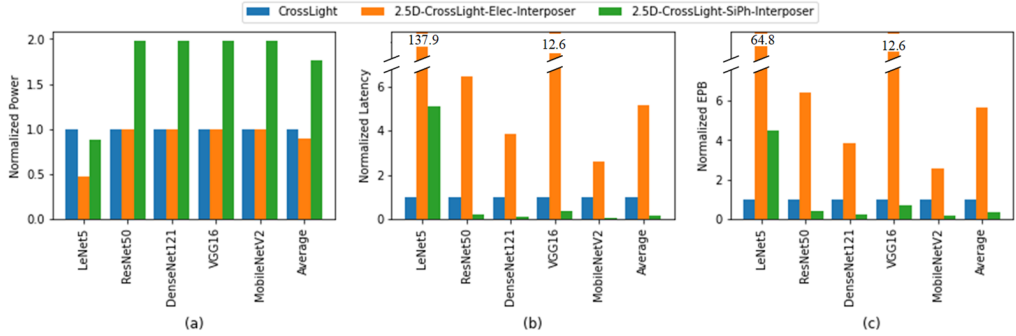

Fig. 6: Performance analysis of CrossLight, 2.5D-CrossLight with electronic interposer, and 2.5D-CrossLight with silicon photonic interposer, (a) normalized power consumption, (b) normalized total latency, and (c) normalized energy-per-bit. From [12].

Figure 6 shows the results of the analysis. It can be observed that *2.5D-CrossLight-SiPh-Interposer* generally achieves better energy efficiency and latency for most CNN models compared to *CrossLight*, except smaller ones like LeNet5 which take up a small fraction of the overall chiplet compute real estate and thus inefficiently utilize the resources on the 2.5D platform. For other models, the performance boost over *CrossLight* is credited to the heterogeneous chiplets and the high-bandwidth photonic interposer network. The *2.5D-CrossLight-Elec-Interposer* variant consumes less power than *2.5D-CrossLight-SiPh-Interposer*, but it faces challenges with higher latency due to metallic interconnects, particularly over longer distances on the large interposers. On average, the *2.5D-CrossLight-SiPh-Interposer* variant demonstrates a significant performance enhancement over the traditional monolithic *CrossLight*, with a 6.6× reduction in latency and a 2.8× decrease in energy-per-bit (EPB). When compared to *2.5D-CrossLight-Elec-Interposer*, it shows even more remarkable improvements: 34× lower latency and 15.8× lower EPB. These improvements are largely due to the *2.5D-CrossLight-SiPh-Interposer*'s capacity to efficiently select specific chiplets and map different CNN model layers to them and adjust inter-chiplet bandwidth as needed. This positions the photonics-based 2.5D chiplet platform as a promising solution for accelerating large-scale ML models.

## VI. Conclusions

In this paper, we discussed how silicon photonics can be leveraged for ML acceleration on a 2.5D chiplet platform that utilizes silicon photonics for both inter-chiplet communication and on-chiplet computation. We discussed the TRINE photonic interposer network architecture which offers substantially lower energy consumption for the same bandwidth of operation when compared to its counterparts. We also discussed how the 2.5D-CrossLight photonic ML accelerator that uses photonic communication and computation can enable low latency and low EPB ML acceleration on chiplet platforms. The 2.5D chiplet platform approach is particularly promising as it allows for the heterogeneous design of chiplets and the integration of off-the-shelf components, enabling the customization of systems to meet various computational needs and capabilities.

## References

[1] AWS Inferentia AI accelerator, Accessed: Feb 1, 2024. [Online]: https://aws.amazon.com/machine-learning/inferentia/
[2] V. Sundaram, et al. "Low cost, high performance, and high reliability 2.5 D silicon interposer," In *IEEE ECTC*, 2013.
[3] J. Kim, et al. "Architecture, chip, and package codesign flow for interposer-based 2.5-D chiplet integration enabling heterogeneous IP reuse." *In IEEE TVLSI*, vol. 28, no. 11, pp: 2424-2437, 2020.
[4] "AMD CDNA3 Architecture", whitepaper, AMD, Accessed: Feb 1, 2024. [Online]:https://www.amd.com/content/dam/amd/en/documents/instinct-tech-docs/white-papers/amd-cdna-3-white-paper.pdf.
[5] P. Fotouhi, *et al.*, "Enabling scalable disintegrated computing systems with AWGR-based 2.5D interconnection networks," *In Journal of Optical Communications and Networking*, vol. 11, no. 7, 2019.
[6] S. V. R. Chittamuru, I. Thakkar, S. Pasricha, "PICO: Mitigating Heterodyne Crosstalk Due to Process Variations and Intermodulation Effects in Photonic NoCs," *In IEEE/ACM DAC 2016*.
[7] D. A. B. Miller, "Device requirements for optical interconnects to silicon chips," *In Proc. IEEE*, vol. 97, no. 7, pp. 1166–1185, 2009.
[8] S. V. R. Chittamuru, S. Desai, S. Pasricha, "SWIFTNoC: A reconfigurable silicon-photonic network with multicast enabled channel sharing for multicore architectures," *In ACM JETC*, 2017.
[9] H. Kwon, *et al.*, "MAERI: Enabling flexible dataflow mapping over DNN accelerators via reconfigurable interconnects," *In ASPLOS*, 2018.
[10] F. Sunny, E. Taheri, M. Nikdast, and S. Pasricha, "A survey on silicon photonics for deep learning," *In ACM JETC*, vol. 17, no. 61, 2021.
[11] E. Taheri, et al. "TRINE: A Tree-Based Silicon Photonic Interposer Network for Energy-Efficient 2.5 D Machine Learning Acceleration." *In IEEE NocArc*, Oct 2023.
[12] F. Sunny, E. Taheri, M. Nikdast, S. Pasricha, "Machine Learning Accelerators in 2.5D Chiplet Platforms with Silicon Photonics," *In IEEE/ACM DATE, 2023*.
[13] E. Taheri, et al. "ReSiPI: A reconfigurable silicon-photonic 2.5 D chiplet network with PCMs for energy-efficient interposer communication." *In IEEE/ACM ICCAD, 2022*.
[14] Y. Li et al. "SPRINT: a high-performance, energy-efficient, and scalable chipletbased accelerator with photonic interconnects for cnn inference", *In IEEE TPDS*, 2021.
[15] Y. Li et al. "SPACX: Silicon photonics-based scalable chiplet accelerator for dnn inference" *In HPCA*, 2022.
[16] F. Sunny, A. Mirza, M. Nikdast, S. Pasricha, "CrossLight: A cross-layer optimized silicon photonic neural network accelerator." *In DAC, 2021*.
[17] S. Afifi, F. Sunny, M. Nikdast, S. Pasricha, "TRON: Transformer Neural Network Acceleration with Non-Coherent Silicon Photonics", *In ACM GLSVLSI, 2023*.
[18] F. Sunny, M. Nikdast and S. Pasricha, "RecLight: A Recurrent Neural Network Accelerator With Integrated Silicon Photonics", *ISVLSI, 2022*.
[19] S. Afifi, F. Sunny, A. Shaifee, M. Nikdast, S. Pasricha, "GHOST: A Graph Neural Network Accelerator using Silicon Photonics", *In ACM Transactions on Embedded Computing Systems (TECS), 2023*.
[20] S. Pasricha, M. Nikdast, "A Survey of Silicon Photonics for Energy Efficient Manycore Computing" *In IEEE Design and Test, Aug 2020*.
[21] E. Taheri, et al., "DeFT: A deadlock-free and fault-tolerant routing algorithm for 2.5D chaplet-based networks," *In IEEE/ACM DATE, 2022*.